\pdfoutput=1
\documentclass[12pt,notitlepage]{article}

\usepackage{times}
\usepackage{mathptmx}

\newcommand{\xx}[1]{\section {#1}}

\providecommand{\spacing}[1]{\renewcommand{\baselinestretch}{#1}\small\normalsize}
\providecommand{\doublespacing}{\spacing{1.5}}
\providecommand{\normalspacing}{\spacing{1.0}}

\newcommand{\eql}[2]{\begin{equation} \label{eq:#1} #2 \end{equation}}
\newcommand{\eqr}[1]{Eq.~(\ref{eq:#1})}

\newcommand{\Figr}[1]{Figure~\protect{\ref{fig:#1}}}

\newcommand{\tabr}[1]{Table~\protect{\ref{tab:#1}}}

\newcommand{\tbox}[1]{\begin{tabular}{c} #1 \end{tabular}}
\newcommand{\ylab}[1]{\tbox{\rotatebox{90}{#1}}}

\usepackage[normalem]{ulem}

\newcommand{\eg}{{\em e.g.}}

\newcommand{\mks}[3][~]{\mbox{$\mathrm{{#2}\mbox{{#1}}{#3}}$}}

\newcommand{\ms}[2][~]{\mks[{#1}]{{#2}}{m \; s^{-1}}}

\renewcommand{\eg}{e.g.}

\newcommand{\z}[1]{\mbox{$ z_{#1} $}}

\usepackage[final]{graphicx}
\usepackage{ametsoc}
\graphicspath{ {/Users/rhoffman/NWP/fig/misc/} }

\begin{document}

\title{Neutral stability height correction for ocean winds}

\author{Ross N. Hoffman\footnotemark[1]}

\footnotetext[1]{\em Contact information:
\upshape
Dr. Ross N. Hoffman,
Atmospheric and Environmental Research, Inc.,
131 Hartwell Avenue,
Lexington, MA 02421-3126
Email: ross.n.hoffman@aer.com.}

\maketitle


\begin{abstract}

Adjusting ocean wind observations to a standard height, usually
\mks{10}{m}, requires the use of a boundary layer model, and knowledge of
the thermodynamical variables.
Height adjustment is complicated by the fact that a necessary
parameter, the roughness height, cannot be given in a closed form
solution.
If only the wind and reporting height are known, the best that can be
done is to assume neutral stability.
The determination of roughness height is analyzed and a simple
approximation \citep[used by][]{AtlHA+11} is derived in detail.
This approximation is accurate for winds in the range of \ms{1-30}
for neutral stratification and would be an excellent initial estimate
for a Newton iteration to determine the roughness height precisely,
whether or not neutral stability is assumed.

\end{abstract}

\xx {Introduction}

Adjusting ocean wind observations to a standard height, usually
\mks{10}{m}, requires the use of a boundary layer model, and knowledge of
the thermodynamical variables.
Whichever PBL model is used, an exact solution requires iterating the
constant flux layer equations.
This is due to the fact that \z0, the roughness length, is an
implicit function of the model variables over the oceans.
The Charnock formula states that over the ocean \z0 and the surface
stress magnitude $|\tau|$ are linearly related by \eql{z0}{\z0 =
\frac{a}{\rho g}|\tau|.}
Here $a = 0.0185$ is the Charnock constant, \mks{g = 9.81}{m \;
s^{-2}}, and $\rho$, the density of air, is assumed to be constant for
the range of heights considered and equal to its surface value.
Previously $a = 0.032$ was the accepted value for the Charnock
constant, but now it is known the Charnock constant is not actually a
constant, but depends on the sea state, through the wave age
\citep{Wu85a}.
Accordingly, the Charnock constant is determined by the wave model in
the ECMWF system, and normally is in the range $0.01 \le a \le 0.04$,
corresponding to sea states from swell to steep young ocean waves
\citep{Her11}.
While values of $a$ as large as 0.1 sometimes occur in the ECMWF
system, a typical value is $a=0.018$, which agrees with the value of
0.0185 used here and by \citet{Wu85a}.
Note that \eqr{z0} neglects the contribution of molecular viscosity
which is important at low wind speeds \citep[\eg,][]{Her11}.
However, at low wind speeds the height correction and consequently
errors made in the height correction should be small.

Surface stress determined from \eql{tau}{\tau = -\rho C_d|V|V } also
depends on \z0 through the neutral drag coefficient, and in the
unstable case through the similarity function, usually denoted $f(Ri)$
where $Ri$ is the Richardson number.
In \eqr{tau} $V$ is the vector wind at some height $z$, $|V|$ is the
magnitude of the vector wind, the wind and stress vectors are assumed
to be parallel for the range of heights considered, and the drag
coefficient $C_d$ is given by the product of the similarity function
$f(Ri)$ and the neutral drag coefficient
\eql{Cdn}{C_{dn} = \left[\frac{k}{\log\left(\frac{\displaystyle z}
{\z0}\right)}\right]^2.}
Here the von K\'{a}rm\'{a}n constant $k = 0.4$.
See \cite{HofL90} for details.

NWP models usually ``cheat'' and use the value of $\tau$ of the
previous time step to find \z0 through the Charnock formula.
Actually using old values to evaluate the dissipative terms can be a
good policy as this can reduce computational instability.
But outside of a model we must calculate \z0 implicitly.
For this purpose we substitute the absolute value of \eqr{tau} into 
\eqr{z0} and then use the expresseion for $C_d$ to obtain
\eql{hz}{\z0 = \frac{a}{g} C_d |V|^2 = \frac{a}{g} f(Ri)
\left[\frac{k|V|}{\log\left(\frac{\displaystyle z}{\z0}\right)}\right]^2 
\equiv h(\z0; |V|, z, Ri).} 
For neutral stratification, $Ri = 0$, and $f(Ri) = 1$.
To solve \eqr{hz} we must iterate.
To begin the process, \cite{HofL90} estimated $C_d$ as a linear
function of $|V|$ and then obtained the initial estimate of \z0 from
the Charnock relationship.
Then \eqr{hz}, $\z0 = h$, is iterated.
This converges to a good approximation within a few iterations.
It is then possible to switch to a Newton iteration to solve $\z0 - h
\equiv f(\z0) = 0$.
The Newton method requires the partial derivative of $h$ with respect
to \z0.
This can be evaluated using the tangent linear code corresponding to the
calculation of $h$ by setting all inputs to zero except for that
corresponding to \z0, which is set to unity.
The advantage of the Newton method is that it iterates to machine
precision in 2-4 steps from a reasonable start.
(With an unreasonable start it can diverge.)
With a solution exact to machine precision one can then skip the
iteration in the adjoint and/or tangent models.
As an alternative within the context of the ECMWF system,
\citet{Her11} describes an accurate fit for $C_{dn}$ and \z0 as
functions of neutral wind speed and the Charnock value, two parameters
available in the interface between the ECMWF atmospheric and wave
models.

\xx {Height correction for ocean winds}

Knowing \z0 is equivalent to knowing the stress, and we can then
solve \eqr{hz} for $|V|$ at any height, $z$, if only we know the
Richardson number.
In particular, \eqr{hz} states that $ C_d^{\frac{1}{2}} |V| $ is conserved as
we vary the height $z$.
However $Ri$ depends on knowing the stratification of the boundary
layer.
In what follows we assume that only the wind and reporting height are
known.
Then the best that can be done is to assume neutral stability, and in the
rest of this treatment $V$ will denote the neutral stability wind.
Now, $ C_{dn}^{\frac{1}{2}} |V| $ is conserved, allowing
us to determine the \mks{10}{m} neutral wind speed $|V_{10}|$
from an observation at some other height according to:
\eql{V10}{|V_{10}| = \left[ \frac{\log(10/\z0)}{\log(z/\z0)} \right] |V|.}
Note that according to \eqr{V10}, the ratio between neutral
stability winds at two levels is entirely determined by $z_0$ and the
two heights.

Once \z0 is determined, the neutral wind,
defined by \eql{U}{\tau = -\rho C_{dn} |V| V, } is easily determined
from knowledge of \z0 alone according to
\eql{V}{V = \left( \frac{g}{a k^2} \right)^{\frac{1}{2}}  \z0^{\frac{1}{2}}
\log\left(\frac{z}{\z0}\right) , }
which is obtained by combining \eqr{U} with \eqr{z0} and making
use of \eqr{Cdn}.
A few sample calculations using \eqr{V} are presented in
\tabr{sample} for heights of 4, 10, and \mks{19.5}{m}.
In \tabr{sample} we see that the variation in \z0 is two orders of
magnitude greater than the variation in wind speed.
Over this range of wind speed, the correction factors for determining
$|V_{10}|$ vary by as much as 5\%.
This variation is the same order of magnitude as the corrections, and
is therefore worth accounting for.

\newcommand{\titem}[3]{\begin{table}[h] \normalspacing 
 \caption{{#2}\label{tab:#1}}
 \centerline{{#3}} \end{table}}

\titem{sample}
 {Sample calculations based on \eqr{V}.
The \z0 values are equal to  $2^{-j}$ but have been multiplied by
10$^6$ for presentation in this table.
The $|V_{10}|$ values are in \ms{} and are calculated using \eqr{V}.
The ratios in columns 4 and 5 are equal to the term in square
brackets in \eqr{V10}.
The last column contains the estimated value of \z0 from \eqr{z0calc}.}
 {\begin{tabular}{|rr|rrr|r|} \hline \hline
 $j$ & $\z0 \times 10^6$ & $|V_{10}|$ & $|V_{10}/V_{4}|$ & $|V_{10}/V_{19.5}|$ &
 $\hat{z}_0 \times 10^6$ \\ \hline
 14 &   61 &  5.4 & 1.08 & 0.947 &   68 \\
 11 &  488 & 12.6 & 1.10 & 0.937 &  493 \\
  8 & 3906 & 28.2 & 1.13 & 0.922 & 3216 \\ \hline
\end{tabular}}

\xx {Calculation of \z0 under neutral conditions}

To apply \eqr{V10} we still need to determine \z0.
Here we demonstrate a simple approximation.
The motivation is that under neutral conditions, for some fixed
height, we expect wind speed, surface stress, and roughness height to
all increase together.
Differentiating \eqr{V}, we obtain
\eql{dVdz0}{\frac{d|V|}{d\z0} = \left(\frac{g}{a k^2}\right)^{\frac{1}{2}} \z0^{-\frac{1}{2}} 
 \left[\frac{1}{2} \log\left(\frac{z}{\z0}\right) - 1\right] . } 
Thus $d|V|/d\z0/ > 0$ provided $z > e^2 \z0$.
This holds for wind speeds less than hurricane strength and heights of
several meters or more.
The suggestion then is that \z0 should be a monotonically increasing function
of $|V|$, and interpolation into a look-up table, or a simple fit to
a set of exact values should work.

For this investigation it is convenient to define
\eql{y}{y = \log(z/\z0) \quad \mbox{so that} \quad \z0 = z e^{-y} . }
Then the square of \eqr{V} may be written as
\eql{gamma}{ y^2 e^{-y} \equiv \gamma = \frac{a k^2}{g z} V^2 . }
Usually we will know $V$ and $z$ and hence $\gamma$ from \eqr{gamma}.
From $\gamma$ we then determine $y$ and finally \z0 from \eqr{y}.
To determine $y$ from $\gamma$ we  tabulate or model $y$
as a function of $\gamma$, based on data obtained by calculating
$\gamma$ from \eqr{gamma} for different values of $y$.
Note that the values of the regression coefficients determined below
are independent of the value of $z$ or any of the other parameters,
including the Charnock constant $a$.
However, to create a relevant sample of $y$-values for fitting, we
take \mks{z=10}{m}, and vary \z0.
Below, as in \tabr{sample}, we take values of \z0 evenly distributed
in log space given by $2^{-j}$ for integer values of $j$.

\newcommand{\fitem}[3]{\begin{figure}[hb] \normalspacing
 \caption{{#2}\label{fig:#1}} \medskip
 \centerline{{\sffamily\upshape\bfseries
 {#3}}} \end{figure}}

\Figr{fit}a plots $y$ as a function of $\log(\gamma)$ for $j =
0, \ldots, 30$.
Clearly a linear fit will work well over most the range of $\gamma$.  
This is not unexpected since according to \eqr{gamma} $\log \gamma = -
y + 2 \log y$.
As $j$ increases, \z0, $|V_{10}|$, and $\gamma$ decrease, while $y$
increases.
For example, for $j = 0$, \mks{\z0 = 1}{m} and \ms{|V_{10}| = 134},
while for $j = 30$, \mks{\z0 \approx 10^{-9}}{m} and \ms{|V_{10}| =
0.04}.
Fitting points for $j \ge 6$, corresponding to \ms{|V_{10}| \le 40}
(where the extreme point included is marked by the vertical line in
the plots) we find that
\eql{ycalc}{y = c_0 + c_1 \log(\gamma)}
with $c_0 = 3.7$ and $c_1 = -1.165$.

Combining \eqr{y}, \eqr{gamma}, and \eqr{ycalc}, we obtain our
estimate of \z0
\eql{z0calc}{\hat{z}_0=z \exp[-(c_0 + c_1 \log[\gamma])]=z \exp[-(c_0 + c_1 \log[(a k^2 V^2)/(g z)])] . }
\Figr{fit}b shows the error of the fit in log space.
Values calculated using \eqr{z0calc} are shown in the last column of
\tabr{sample} for the cases listed.
The differences are not tiny, but when we recalculate the ratios of
the wind speeds in columns 4 and 5 of the table, the results are
nearly the same.
Using the same precision as in the table, the values are the same
except that the value for $|V_{10}/V_{19.5}|$ is 0.923.

\clearpage

\fitem{fit} {Fitting $y = \log(z/\z0)$ as a function of $\gamma = (a
k^2/g z) V^2$.
(a) The linear fit to values of $y$ and $\log \gamma$ for $j = 6, \ldots,
30$ is plotted as a dotted line.  The data values are plotted as
dots.
(b) Log residuals for \eqr{z0calc} fit of \z0.  For this calculation
we define the true values of \z0 and then $y$, $\gamma$ and $\hat{z}_0$ for
the chosen value of $z$ (\mks{10}{m}), using \eqr{y} and \eqr{z0calc}.
The vertical lines identify $j=6$.}
 {\setlength{\tabcolsep}{1pt}
 \begin{tabular}{ccccc}
 \ylab{$y$} & \tbox{\includegraphics[scale=.5, viewport=115 200 500 585,
 clip=true]{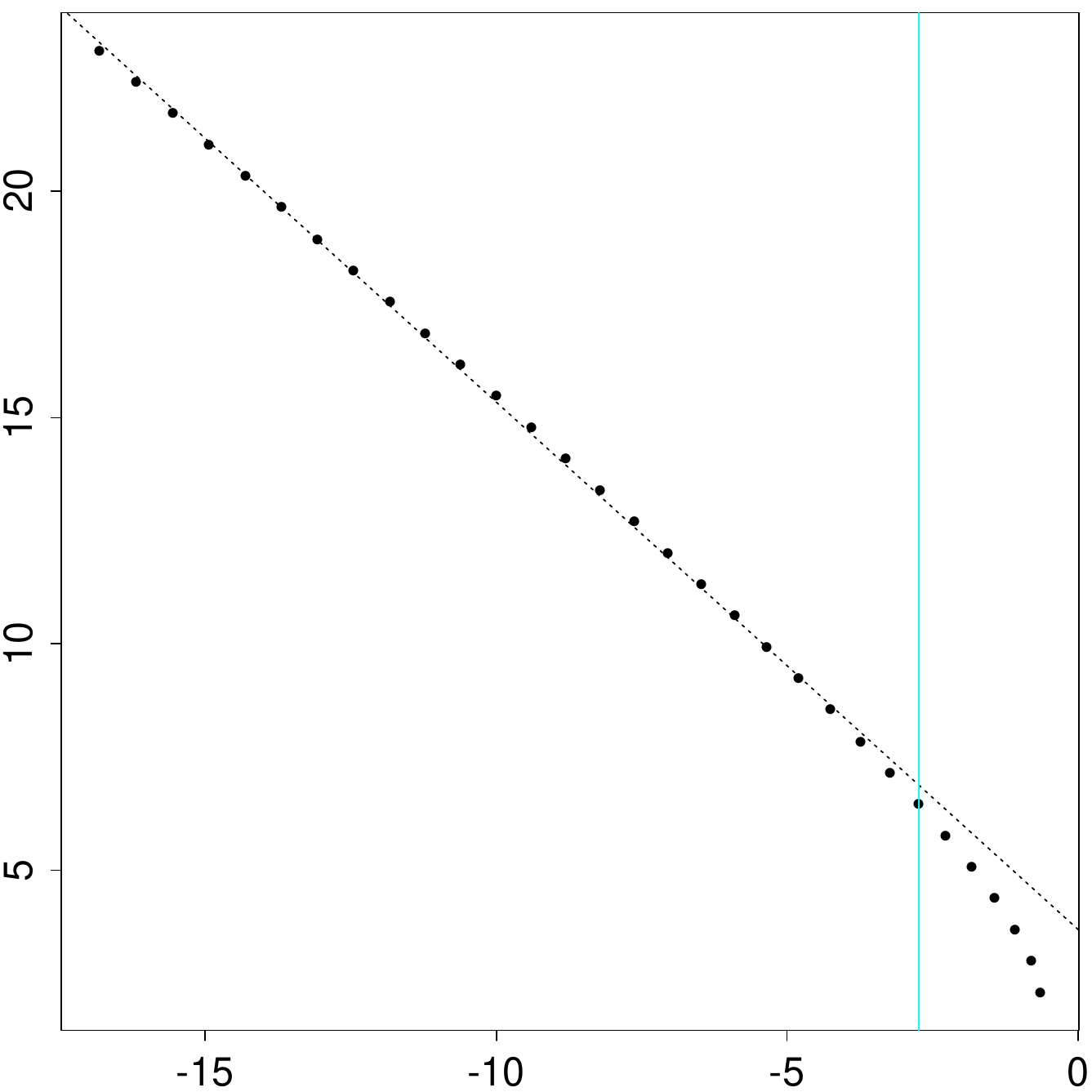}} & &
 \ylab{$\log \hat{z}_0 - \log \z0$} & \tbox{\includegraphics[scale=.5,
 viewport=115 200 500 585, clip=true]{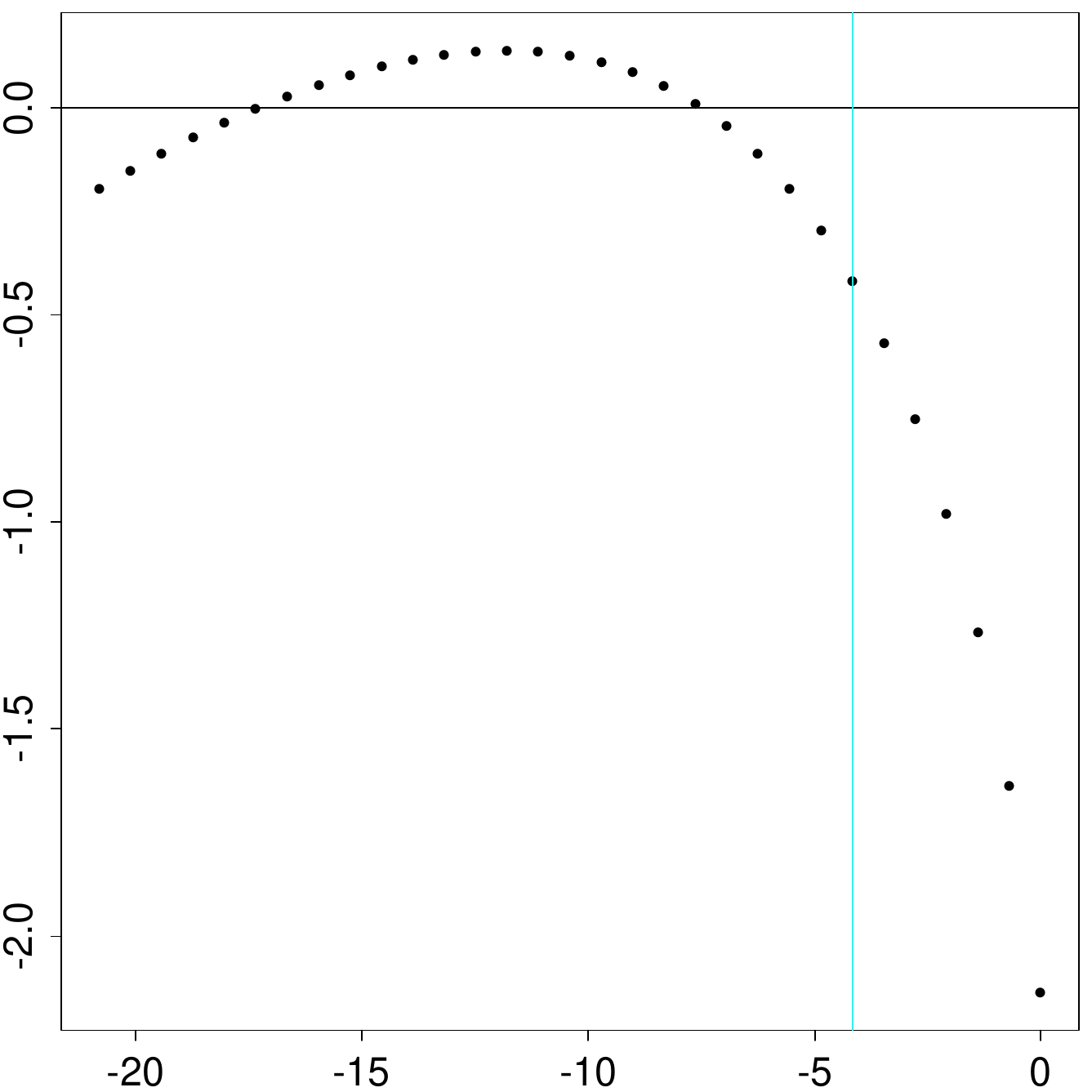}} \\
 a & $\log \gamma$ & \makebox[15mm]{~} & b & $\log \z0$ 
 \end{tabular}}

\xx {Concluding remarks}

An approximation for roughness height (\eqr{z0calc}) is given that is
accurate for winds in the range of \ms{1-30}, assuming neutral
stability and neglecting effects of molecular viscosity.
(Also, note that coherent structures in the atmospheric boundary layer
are not explicitly included in the similarity theory employed here.)
Values of \z0 that vary with wind speed should be used in correcting
ocean winds to a standard height.
Typical corrections are in the range of 5-15\%, so assuming a single
value for \z0 will incur errors of a few per cent.
Our approximation is also an excellent initial estimate to
begin a Newton iteration to determine the roughness height precisely,
whether or not neutral stability is assumed.

In practice, the approximation derived here is adequate because errors
due to other approximations and assumptions are graver.
For example, observations and meta-data associated with ship reports
are often limited: information required to estimate atmospheric
boundary layer stability may be lacking, anemometer heights may be
unknown or incorrect, and effects due to ship motion and flow
anomalies due to superstructure may not be accounted for.

\clearpage

\acknowledgement{
The NASA Ocean Surface Wind Projects and the NASA Earth System Data
Records Programs supported this work.
Juan Carlos Jusem (Goddard Space Flight Center) and Zongpei Jiang
(National Oceanography Centre, Southampton) provided helpful comments
on the manuscript.
A preliminary version of this study was posted to: \\
http://map.nasa.gov/data/ssw.old/reason\_sample/doc/src/height-correct.pdf}


\normalspacing


\end{document}